\documentstyle[multicol,epsfig,prb,aps]{revtex}

\begin{document}


\title{Structure determination of the indium-induced Si(111)-(4$\times$1) 
  reconstruction by surface x-ray diffraction}

\author{O. Bunk, \cite{author_email} 
  G. Falkenberg, J.H. Zeysing, L. Lottermoser, and R.L. Johnson} 
\address{II. Institut f\"ur
  Experimentalphysik, Universit\"at Hamburg, Luruper Chaussee 149,\\
  D-22761 Hamburg, Germany}

\author{M. Nielsen, F. Berg-Rasmussen, J. Baker, and R. Feidenhans'l}
\address{Condensed Matter Physics and Chemistry Department,\\
  Ris\o\ National Laboratory, DK-4000 Roskilde, Denmark}

\date{21 October 1998}

\maketitle

\begin{abstract}
A detailed structural model for the indium-induced 
Si(111)-(4$\times$1) surface reconstruction has been 
determined by analyzing an extensive set of x-ray-diffraction data recorded 
with monochromatic ($\hbar\omega$ = 9.1~keV) synchrotron radiation. 
The reconstruction is quasi-one-dimensional. The main features 
in the structure are chains of silicon atoms alternating with 
zigzag chains of indium atoms on top of an essentially 
unperturbed silicon lattice. 
The indium coverage corresponds to one monolayer. 
The structural model consistently explains all previously published 
experimental data. 
\end{abstract}

\pacs{
PACS: 68.35.Bs    
}


\begin{multicols}{2}


Considerable interest has focused recently on ad\-sor\-bate-induced 
modification of semiconductor surfaces as a technique to create 
nanoscale quantum structures of high perfection. 
In this paper we report the formation of quasi-one-dimensional (1d) chains 
on the (4$\times$1)-reconstructed Si(111) surface induced by the adsorption 
of indium and present the atomic structure as determined by 
surface x-ray diffraction (SXRD). 

Several surface reconstructions are induced by indium 
on Si(111) (see, e.g., Kraft {\it et al.}, 
Ref.\ \onlinecite{kraft_si111in_review}) depending on the coverage; 
the Si(111)-(4$\times$1)-In 
reconstruction marks the borderline between the semiconducting, low indium 
coverage, and metallic,  high-coverage, phases. \cite{kraft_si111in_review} 
Despite the fact that the indium-induced Si(111)-(4$\times$1) 
reconstruction was reported for the first time by Lander and Morrison 
in 1965\cite{landermorrison_si111alin} and has been investigated with 
a variety of techniques subsequently, no definitive structural model could be 
established to date and some of the experimental results are apparently 
inconsistent. 

In direct \cite{abukawa_si1114x1in_arpes} and inverse 
\cite{hillmclean_si1114x1in_ipes_prb56} photoemission investigations on 
single-domain samples and recently in inverse photoemission 
investigations on a three-domain sample \cite{hillmclean_si1114x1in_3dom_ipes} 
the Si(111)-(4$\times$1)-In reconstruction showed a quasi-1d metallic 
behavior. 
Additionally to this interesting electronic property the surface exhibits 
an image state that also showed quasi-1d behavior in its strongly 
anisotropic dispersion: Along a certain direction the dispersion is very well 
described by a free electron parabola whereas in the perpendicular direction 
the dispersion falls below the free electron parabola. 
\cite{hillmclean_ipes_imagestate} 
Scanning tunneling microscopy (STM) investigations of the 
Si(111)-(4$\times$1)-In reconstruction 
\cite{kraft_si111in_review,nogami_si111in_prb36,stevens_si1114x1in_prb47,%
saranin_si1114x1in_prb56,saranin_si1114x1in_prb55,%
owmanmartensson_si111s3xs3in_stm}
resolved zig-zag chains running in the $\langle\overline{1}10\rangle$ 
directions in the filled-state images and linear chains in the 
empty-state images. Tunneling data was acquired at bias 
voltages down to 0.04~V consistent with metallic behavior in agreement 
with scanning tunneling spectroscopy results. \cite{kraft_si111in_review} 
Adsorbed hydrogen was found to displace the indium atoms. 
\cite{saranin_si1114x1in_prb56,saranin_si1114x1in_prb55,%
owmanmartensson_si111s3xs3in_stm} 
Filled-state STM images of the hydrogenated substrate 
show a (4$\times$1) reconstruction with straight chains 
instead of the broad zigzag chains on the indium-terminated surface. 
\cite{saranin_si1114x1in_prb56,saranin_si1114x1in_prb55,%
owmanmartensson_si111s3xs3in_stm} 
From these observations and results found in the literature 
Saranin {\it et al.} \cite{saranin_si1114x1in_prb56} 
proposed a structural model. 
This model as well as the model derived by Collazo-Davila {\it et al.} 
\cite{ldm_si1114x1in_surfrevlett4} by applying 
direct methods to transmission electron diffraction (TED) data are at 
variance with the x-ray photoelectron spectroscopy (XPS) results of 
Abukawa {\it et al.} \cite{abukawa_si1114x1in_xps} 
Since the core-level spectra did not show peaks corresponding to surface 
silicon atoms a complex substrate reconstruction can be ruled out. 

All the inconsistencies between the earlier experimental 
results are eliminated with our structural model. 
The determination of the geometric structure lays the foundation for a 
theoretical investigation of the interesting electronic 
structure of this system. 


We used the well established experimental technique SXRD to determine 
the structure of the Si(111)-(4$\times$1)-In reconstruction. 
To minimize the uncertainties induced by not well controlled sample 
preparation conditions we employed the unique combination of an 
ultra high vacuum (UHV) system equipped with standard techniques like 
reflection high energy electron diffraction and 
low energy electron diffraction (LEED) facilities close to the wiggler 
beamline BW2 at HASYLAB (Hamburger Synchrotronstrahlungslabor) 
and a portable UHV chamber for the SXRD measurements. 

We used p-type (B doped) Si(111) wafers (Wacker Chemie) with a resistivity of 
7~$\Omega$cm and nominally no miscut. 
STM measurements on the clean Si(111)-(7$\times$7) surface showed domain 
sizes of more than 10000~\AA{} so the miscut was less than 0.02$^\circ$. 
The carefully outgased sample was cleaned repeatedly by `flashing' to 
$\sim$1150$^\circ$C for 15-20~s and slow cooling from 900$^\circ$C to 
room temperature. 
Indium was deposited from a Knudsen cell at a rate of 
$\sim$0.4~ML/min on the sample at $\sim$500$^\circ$C until the 
($\sqrt[]{3}\times\sqrt[]{3}$)R30$^\circ$ reconstruction was observed; 
at this stage the stacking fault of the Si(111)-(7$\times$7) reconstruction 
was removed. 
Further deposition at a lower temperature of $\sim$430$^{\circ}$C 
(with a lower probability for indium desorption) yielded first the 
($\sqrt[]{31}\times\sqrt[]{31}$) and finally the ($4\times$1) reconstruction. 
The sample was transferred in a portable UHV chamber to the wiggler 
beamline BW2 at HASYLAB for the x-ray diffraction measurements. 
The incident monochromized x-rays 
with an energy of 9.1~keV impinged on the sample at a grazing angle 
of 0.5$^\circ$. An in-plane data-set of 296 reflections with 
$l=0.07$ was recorded by rotating the 
crystal about the surface normal. The background subtracted integrated 
intensities were corrected for the Lorentz factor, polarization factor, 
active sample area and the rod interception appropriate for the 
z-axis geometry. \cite{vlieg_correction_factors} 
In SXRD measurements the fractional-order reflections which belong to one 
domain do not overlap with the reflections which belong to the other two 
rotational domains. 
By comparing equivalent fractional-order reflections the areas of the 
three rotational domains were found to correspond to 
25\%, 37\% and 38\% of the total surface area. The equivalent 
rescaled fractional-order reflections were averaged under the 
assumption of mirror-lines running along $\langle 11\overline{2}\rangle$ 
and a systematic error of $\epsilon=13.7$\% was determined. 
The rods were scaled according to 
the corresponding in-plane intensity and therefore only one overall 
scale factor was necessary for the data analysis. 
In total the data-set consists of 550 symmetry inequivalent reflections, 
61 along two crystal truncation rods (ctrs), 337 along 11 fractional-order 
rods and 152 in-plane reflections. 

In the following we use LEED coordinates with 
${\bf a} = 1/2 [10\overline{1}]_{\mbox{cubic}}$, 
${\bf b} = 1/2 [\overline{1}10]_{\mbox{cubic}}$ and
${\bf c} = 1/3 [111]_{\mbox{cubic}}$. The cubic coordinates are 
in units of the silicon lattice constant (5.43~\AA) and therefore 
$|a| = |b| = 3.84$~\AA{} and $|c| = 3.14$~\AA. The absolute values 
of the reciprocal coordinates including a factor of $2\pi$ are 
$|a^\ast| = |b^\ast| = 1.89$~\AA$^{-1}$ and 
$|c^\ast| = 2.00$~\AA$^{-1}$.


Usually the first step in the analysis of surface diffraction data is 
to plot the Patterson function, i.e.\ the pair correlation 
function of the electron density. The Patterson function 
projected in the surface plane can be calculated from the 
in-plane reflections which have a small momentum transfer perpendicular 
to the surface. A contour plot of this function is shown in 
Fig.~\ref{fig:pat_proj} and is qualitatively in good agreement with that 
previously reported by Finney {\it et al.} \cite{finney_si1114x1in_sxrd} 
Each peak in the Patterson function corresponds to an important interatomic 
vector in the surface reconstruction. Since indium ($Z^2=49^2$) is a much 
stronger scatterer than silicon ($Z^2=14^2$) the peaks correspond 
to interatomic vectors between indium atoms.  
The three interatomic distance vectors which can be seen in the 
Patterson function in Fig.~\ref{fig:pat_proj} indicate that at least three 
different indium sites must be involved in the reconstruction. 
It is highly unlikely that the indium coverage can be 
0.5~ML (monolayer) since that would only correspond to two indium atoms 
per unit cell. 
It has been argued that a reconstruction of 
the substrate might lead to further peaks in the Patterson function 
\cite{ldm_si1114x1in_surfrevlett4} but the XPS data 
\cite{abukawa_si1114x1in_xps} make this unlikely. 

During the course of the data analysis we tested 
several possible initial indium atom configurations and determined 
the substrate structure in the subsequent refinement. 
The model found is shown in Fig.~\ref{fig:model}. 
The correctness of the model is proven both by the good agreement of the 
measured data with the intensities calculated from the model structure 
shown in Fig.~\ref{fig:data} and 
by the overall reduced $\chi^2$ value of 1.5. We will now describe the 
building blocks of the model and resolve some of the apparent 
inconsistencies between the previously reported experimental results. 

The model shown in Fig.~\ref{fig:model} consists 
of a zigzag chain of silicon atoms as found in the $\pi$-bonded chain model 
for the (2$\times$1) reconstruction of the Si(111) surface 
\cite{pandey_si1112x1} 
on top of an essentially unreconstructed substrate. This model is consistent 
with the XPS investigations which showed that 
no strong silicon surface component was present in the Si $2p$ spectra. 
\cite{abukawa_si1114x1in_xps} 
Even in the absence of indium atoms the silicon chains on the substrate 
possess a (4$\times$1) periodicity and it is highly likely that the chains 
observed after hydrogen adsorption 
\cite{saranin_si1114x1in_prb56,owmanmartensson_si111s3xs3in_stm}
are made up of silicon atoms. 
The Si(111)-(4$\times$1)-In reconstruction is completed by adding two 
zigzag rows of indium atoms in the space between 
the silicon chains. The arrangement of the indium atoms is similar to the 
arrangement proposed on the basis of $\mu$-probe Auger electron 
diffraction investigations, \cite{nakamura_si1114x1in_aed_surfsci256} 
and also used in a previous SXRD study. 
\cite{finney_si1114x1in_sxrd} 
There are two inequivalent types of indium atoms in 
agreement with the XPS results. \cite{abukawa_si1114x1in_xps} 
The indium atoms next to 
the silicon chains are probably covalently bonded to the silicon chain atoms. 
For the inner indium atoms the bonding configuration is not so obvious. 
In Fig.~\ref{fig:model} a bonding configuration of covalent $p_{x,y,z}$ 
bonds to the neighboring indium atoms and down to the silicon substrate 
under an angle of approximately 90$^\circ$ are 
shown. The nearest neighbor distance of the indium atoms is within 
the range of 2.98~\AA{} to 3.14~\AA{}. 
Electron counting can not be strictly applied to determine the bonding 
configuration because the surface has 1d metallic character. 
If we assume that the bonding configuration shown in Fig.~\ref{fig:model} 
is a first approximation to the more complex real configuration then 
it is evident that the surface free energy of this reconstruction is 
lower than for other models since there are no silicon dangling bonds and all 
the group III indium atoms are trivalently bonded. 
The present SXRD data do not permit an accurate determination of the 
bond charge densities; the detailed bonding configuration will have to be the 
subject of a future theoretical investigation. 

An important question is how the model shown in Fig.~\ref{fig:model} 
with four indium atoms per unit cell (1~ML) can be used to explain the 
results of the TED \cite{ldm_si1114x1in_surfrevlett4} and 
impact collision ion scattering spectrometry (ICISS) investigations 
\cite{stevens_si1114x1in_prb47} 
which resulted in models with only two indium atoms per unit cell. 
In the TED study direct methods were used to analyze the data and 
electron density maps with indium atom configurations compatible with the 
TED data were presented including 
plausible configurations with four indium atoms 
(e.g.\ Fig.~2 (f) in Ref.\ \onlinecite{ldm_si1114x1in_surfrevlett4}). 
The 1/2~ML model which was compatible with the TED 
data is at variance with the SXRD data as shown in Fig.~\ref{fig:data}~(c) 
where the dashed lines calculated using the 
1/2~ML model do not reproduce the measured data adequately. 
A preliminary data analysis of the SXRD data using direct methods resulted 
in the indium atom configuration shown in Fig.~\ref{fig:model} thereby 
indicating the correctness of our model. \cite{ldmarks_privcom} The 
TED data were not sufficient to include dynamical diffraction effects 
and the resolution was probably not good enough to rule out the wrong 
indium atom configuration. \cite{ldmarks_privcom} 
The reduced $\chi^2$ value for the model proposed by Saranin {\it et al.} 
\cite{saranin_si1114x1in_prb56} is 7.3 and therefore this model 
can also be definitely eliminated. 
The side view of the Si(111)-(4$\times$1)-In 
reconstruction shown in Fig.~\ref{fig:model}(b) and the atomic coordinates 
given in Table \ref{tab:pos} show that the atoms in the topmost layer 
have three different heights. 
The highest atoms are the indium atoms 
bonded to the silicon chain, the inner indium atoms of the indium ``stripe'' 
are lower and the silicon chain atoms are lowest. The existence of different 
heights is in agreement with the results of a previously performed 
STM investigation. \cite{owmanmartensson_si111s3xs3in_stm} 
Low energy ion scattering is very sensitive to the topmost surface layer. 
The configuration of the topmost indium atoms bonded to the silicon chains 
strongly resembles the 1/2~ML model with indium atoms on 
$H_3$ and $T_4$ sites that has been found to reproduce the ICISS data 
collected using 2~keV Li$^+$ ions. \cite{stevens_si1114x1in_prb47} 
Another 1/2~ML model 
with a zigzag chain of indium atoms on $T_4$ sites also gave reasonable 
agreement with the ICISS data \cite{stevens_si1114x1in_prb47} 
since such chains are present in our model as shown in Fig.~\ref{fig:model}. 


In summary, the structure of the Si(111)-(4$\times$1)-In reconstruction 
has been determined  using surface x-ray diffraction. 
The quasi-one-dimensional 
character of this surface reconstruction is given by zigzag chains of 
silicon atoms on top of an unreconstructed silicon substrate and four indium 
atoms per unit cell (1~ML) arranged in two zigzag chains in the gap 
between the silicon chains. 
The indium atom arrangement may also be regarded as being 
quasi hexagonal. We have shown that most of the previously published 
experimental data are consistent with the new structural model. 
We hope that the atomic coordinates given in Table \ref{tab:pos} will pave 
the way for detailed theoretical investigations of the interesting electronic 
structure of this system. 


We thank the staff of HASYLAB for their technical assistance. 
Financial support from the Danish Research Council through Dansync, 
the Bundesministerium f\"ur Bildung, Wissenschaft, 
Forschung und Technologie (BMBF) under project no. 05622GUA1 and the 
Volks\-wagen Stiftung is gratefully acknowledged.


\newpage


\begin{table}[h]
  \begin{center}
    \begin{tabular}[h]{|c|c|c|d|}
      & position [LEED-coord.]  & deviation $\bf d$ [\AA] %
      &$|{\bf d}|$ [\AA] \\\hline\hline
      In & (0.11,0.06, 0.86) &                     &      \\\hline
      In & (0.86,0.93, 0.85) &                     &      \\\hline
      In & (1.53,0.77, 0.99) &                     &      \\\hline
      In & (3.43,0.22, 0.99) &                     &      \\\hline
      Si & (2.28,0.14, 0.73) &                     &      \\\hline
      Si & (2.71,0.86, 0.76) &                     &      \\\hline\hline
      Si & (0.31,0.65,-0.26) & (-0.11,-0.05,-0.03) &  0.10\\\hline
      Si & (3.96,0.98,-0.00) & (-0.15,-0.08,-0.01) &  0.13\\\hline
      Si & (1.30,0.65,-0.25) & (-0.12,-0.06, 0.01) &  0.10\\\hline
      Si & (0.96,0.98, 0.04) & (-0.16,-0.08, 0.13) &  0.19\\\hline
      Si & (2.29,0.64,-0.33) & (-0.17,-0.09,-0.24) &  0.28\\\hline
      Si & (2.01,0.00,-0.03) & ( 0.03, 0.02,-0.10) &  0.10\\\hline
      Si & (3.29,0.65,-0.23) & (-0.16,-0.08, 0.05) &  0.15\\\hline
      Si & (2.95,0.97,-0.01) & (-0.20,-0.10,-0.03) &  0.18\\\hline\hline
      Si & (0.66,0.33,-1.24) & (-0.04,-0.02, 0.04) &  0.07\\\hline
      Si & (0.33,0.67,-1.01) & (-0.01,-0.01,-0.02) &  0.02\\\hline
      Si & (1.65,0.32,-1.26) & (-0.08,-0.04,-0.03) &  0.07\\\hline
      Si & (1.32,0.66,-0.99) & (-0.06,-0.03, 0.03) &  0.06\\\hline
      Si & (2.67,0.33,-1.27) & ( 0.00, 0.00,-0.05) &  0.05\\\hline
      Si & (2.32,0.66,-1.04) & (-0.05,-0.03,-0.13) &  0.14\\\hline
      Si & (3.66,0.33,-1.25) & (-0.04,-0.02, 0.00) &  0.03\\\hline
      Si & (3.31,0.65,-0.98) & (-0.09,-0.05, 0.04) &  0.09\\\hline
    \end{tabular}
  \end{center}
  \caption{The atom positions in the Si(111)-(4$\times$1)-In reconstruction 
    derived from the analysis of the SXRD data. The positions of the atoms 
    are given 
    in LEED coordinates, the deviations from the bulk-like positions and 
    the absolute values of these deviations are given in \AA{}. For the 
    silicon atoms an isotropic Debye-Waller factor of 0.5~\AA$^2$ and for the 
    indium atoms of 4.5~\AA$^2$ was used in the data analysis.}
  \label{tab:pos}
\end{table}


\begin{figure}[h]
  \caption{Patterson function of the electron density projected 
    in the surface plane calculated from the fractional-order in-plane 
    reflections. 
    The axes are scaled in LEED coordinates (1.0 corresponds to 3.84~\AA).
    The distance vectors shown are 1. (0.69,0.85), 2. (1.42,0.71) and 
    3. (2.00,0.50). The dashed line from (0,0) to (2,1) indicates the 
    mirror line (along $\langle 11\overline{2}\rangle$ in bulk coordinates).}
  \label{fig:pat_proj}
\end{figure}

\begin{figure}[h]
  \caption{Ball and stick model of the Si(111)-(4$\times$1)-In 
    reconstruction in top (a) and side (b) views. Indium atoms are 
    drawn dark grey, silicon atoms are drawn light grey. The contributions 
    to the peaks of the Patterson function are shown by arrows. The 
    standard LEED (4$\times$1) unit cell is indicated by a dashed line. 
    The dashed-dotted line along $[11\overline{2}]$ indicates a mirror line.}
  \label{fig:model}
\end{figure}

\begin{figure}[h]
  \caption{Measured and calculated SXRD intensities: 
    (a) in-plane with $l=0.07$. The radii of the filled (empty) semi-circles 
    are proportional to the measured (calculated) intensities. 
    Hatched circles are scaled with a factor of 0.5. 
    (b) Fractional-order rod-scans. The solid line is calculated using 
    the model shown in Fig.~\ref{fig:model} with the coordinates 
    given in Table \ref{tab:pos}.
    (c) Integer order rods. The dashed lines correspond to the 
    intensities calculated using the best fit to the model proposed in 
    Ref.\ \protect\onlinecite{ldm_si1114x1in_surfrevlett4}.}
  \label{fig:data}
\end{figure}
\newpage



~\vspace{2cm}

\begin{center}
  \epsfxsize=3.5cm
  \epsffile{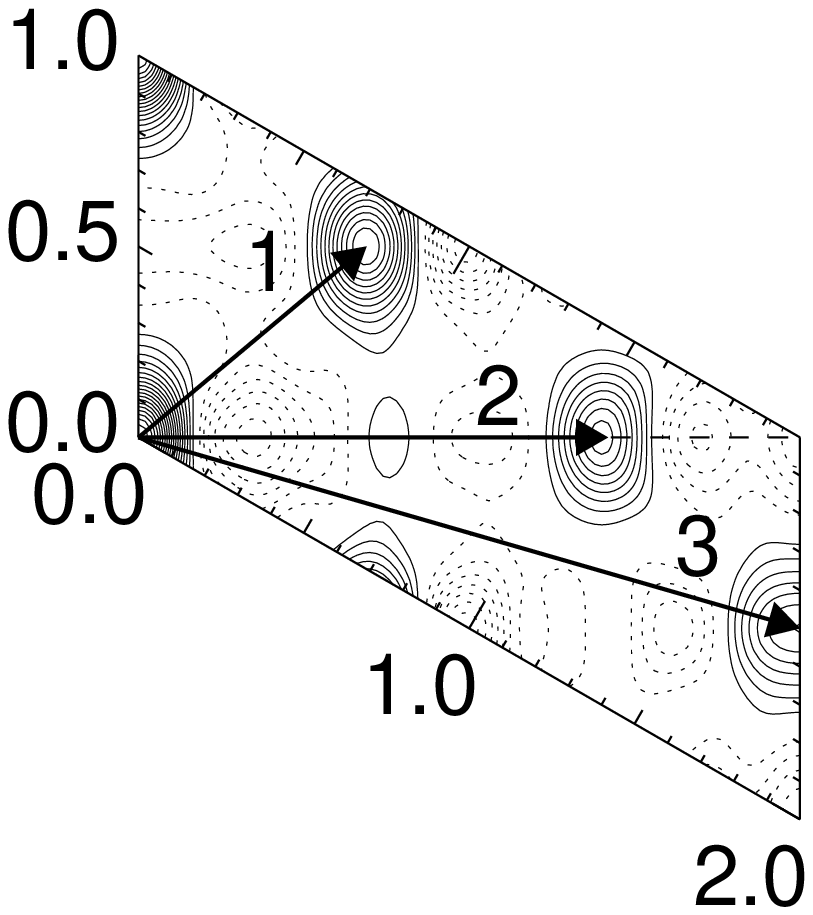}
  \vfill

  {\LARGE Fig.~\ref{fig:pat_proj}}
\end{center}
\vfill


\begin{center}
  ~\vspace{2cm}

  \epsfxsize=5cm
  \epsffile{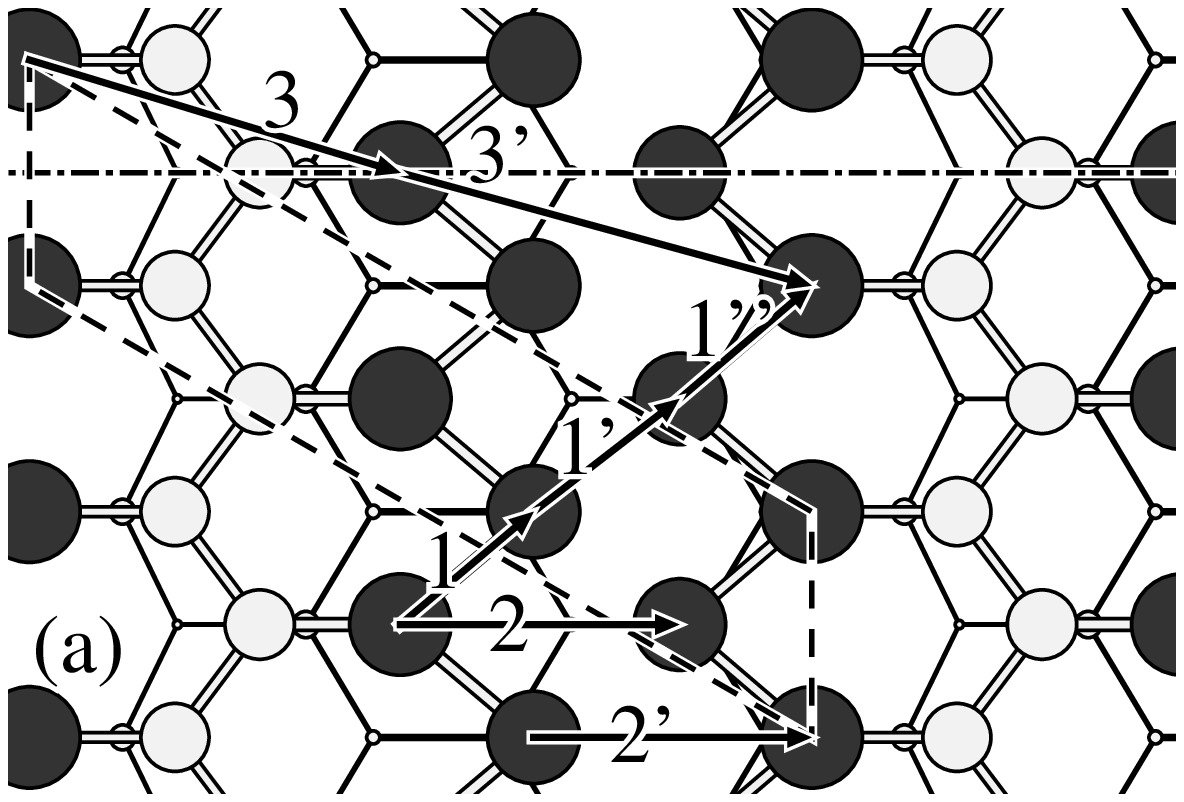}
  \bigskip

  \epsfxsize=5cm
  \epsffile{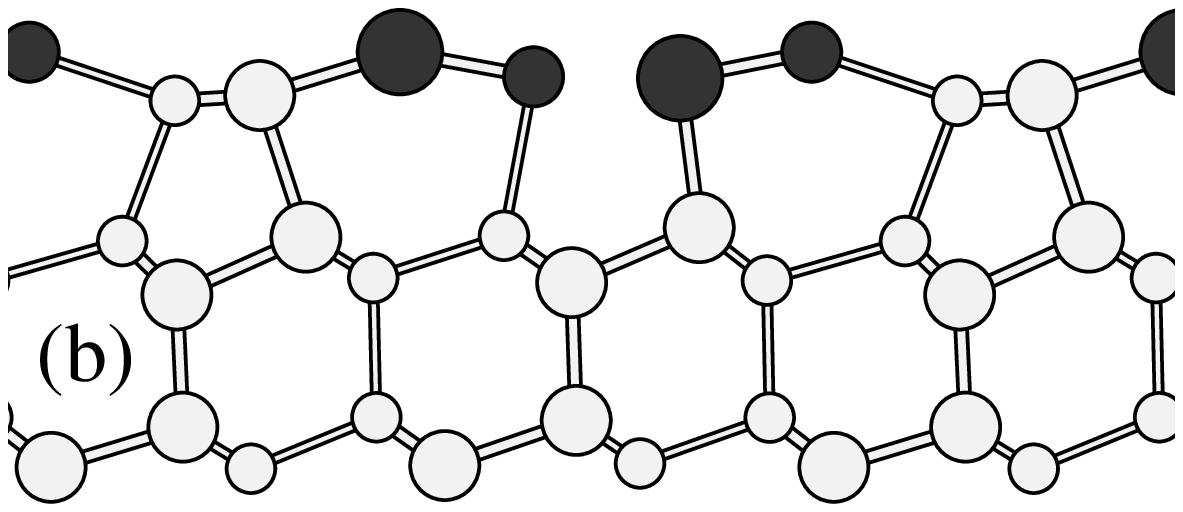}
  \vfill

  {\LARGE Fig.~\ref{fig:model}}
\end{center}
\vfill


\begin{center}
  ~\vspace{2cm}

  \epsfxsize=8.4cm
  \epsffile{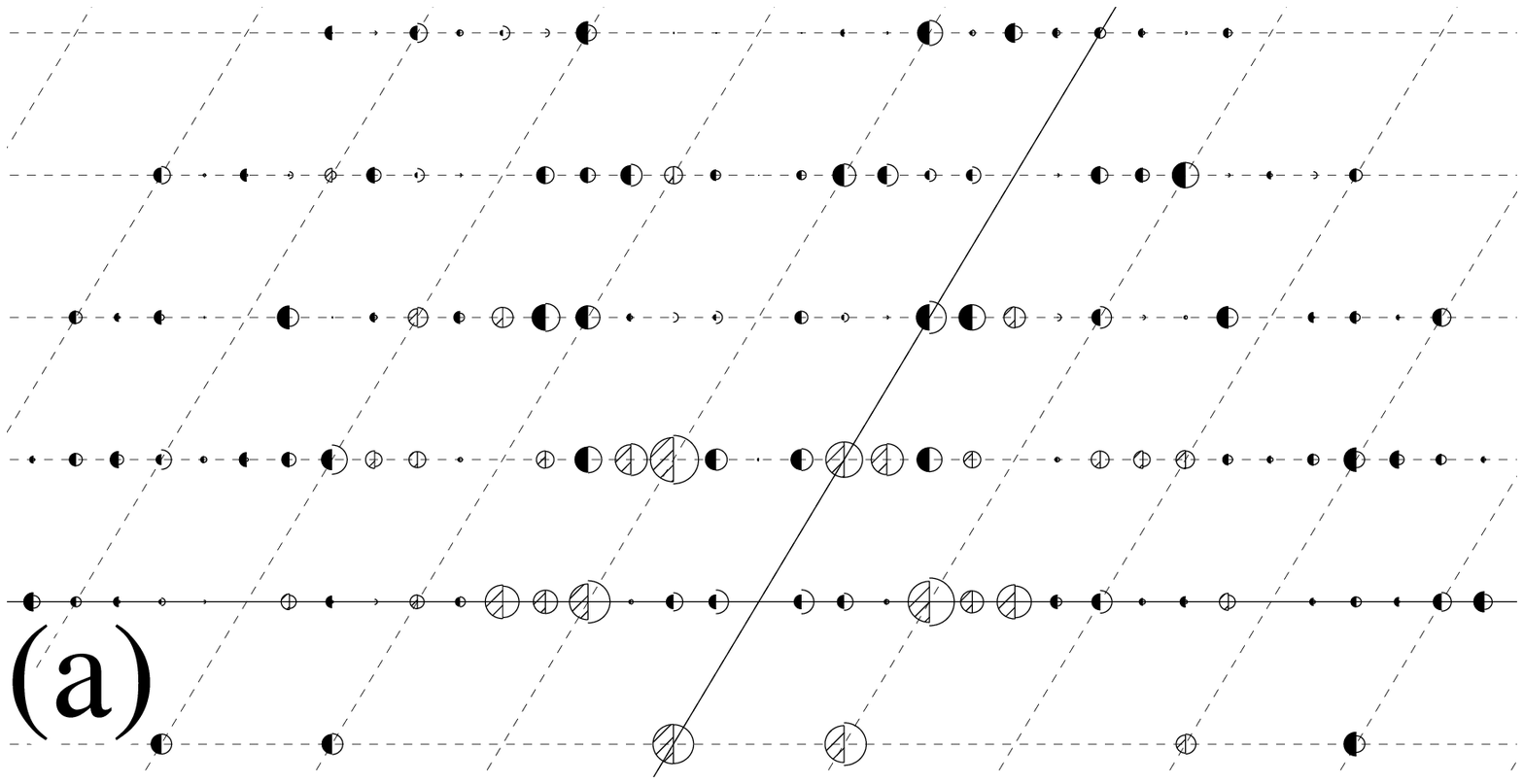}
  \medskip

  \epsfxsize=8.4cm
  \epsffile{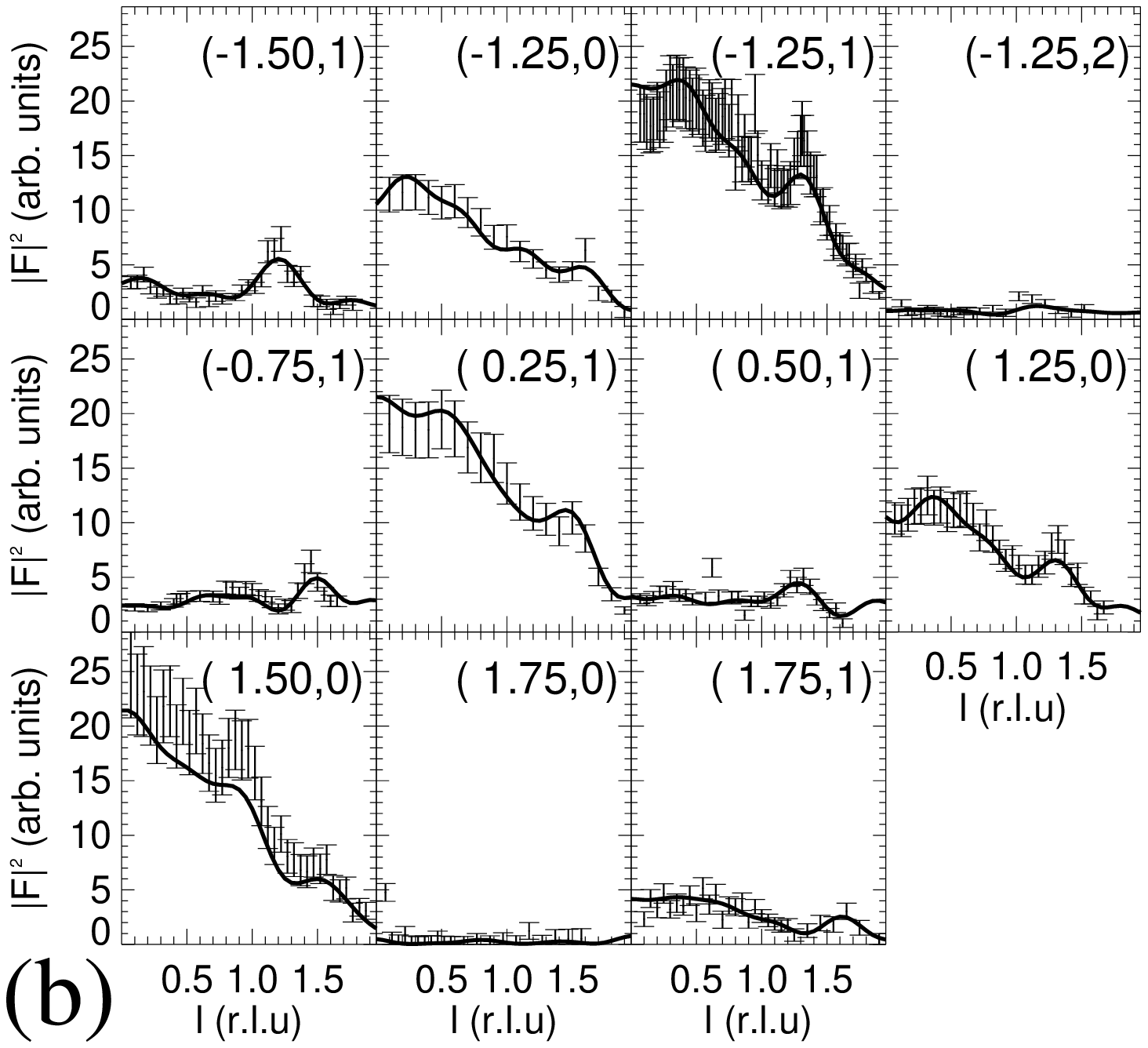}
  \medskip

  \epsfxsize=8.4cm
  \epsffile{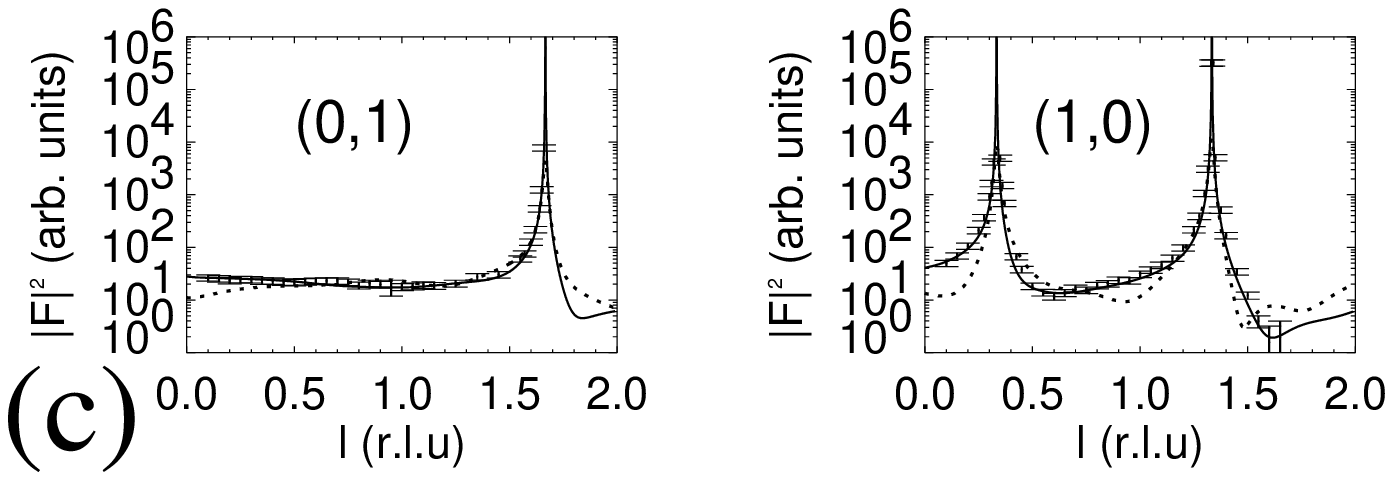}
  \vfill

  {\LARGE Fig.~\ref{fig:data}}
\end{center}
\end{multicols}
\end{document}